\documentclass[10pt]{article} 
\usepackage[preprint]{tmlr}

\usepackage{amsmath,amsfonts,bm}

\def\eqref#1{equation~\ref{#1}}

\def\1{\bm{1}}

\DeclareMathAlphabet{\mathsfit}{\encodingdefault}{\sfdefault}{m}{sl}
\SetMathAlphabet{\mathsfit}{bold}{\encodingdefault}{\sfdefault}{bx}{n}

\usepackage{hyperref}
\usepackage{url}
\usepackage{braket}
\usepackage{graphicx} 
\usepackage{float}
\usepackage{geometry}
\usepackage{caption}
\usepackage{subcaption}
\usepackage{biblatex}
\title{Impact of Data Augmentation on QCNNs }

\author{\name Leting Zhouli \email lzhouli@student.unimelb.edu.au \\
      \addr Faculty of Engineering and Information Technology\\
      The University of Melbourne\\
      \name Peiyong Wang \email peiyongw@student.unimelb.edu.au \\
      \addr Faculty of Engineering and Information Technology\\
      The University of Melbourne\\
      \name Udaya Parampalli \email udaya@unimelb.edu.au \\
      \addr Faculty of Engineering and Information Technology\\
      The University of Melbourne\\
      }

\def\month{MM}  
\def\year{YYYY} 
\def\openreview{\url{https://openreview.net/forum?id=XXXX}} 

\begin{document}

\maketitle

\begin{abstract}
In recent years, Classical Convolutional Neural Networks (CNNs) have been applied for image recognition successfully. Quantum Convolutional Neural Networks (QCNNs) are proposed as a novel generalization to CNNs by using quantum mechanisms. The quantum mechanisms lead to an efficient training process in QCNNs by reducing the size of input from $N$ to $log_2N$. This paper implements and compares both CNNs and QCNNs by testing losses and prediction accuracy on three commonly used datasets. The datasets include the MNIST hand-written digits, Fashion MNIST and cat/dog face images. Additionally, data augmentation (DA), a technique commonly used in CNNs to improve the performance of classification by generating similar images based on original inputs, is also implemented in QCNNs. Surprisingly, the results showed that data augmentation didn’t improve QCNNs performance. The reasons and logic behind this result are discussed, hoping to expand our understanding of Quantum machine learning theory.\\

\end{abstract}

\section{Introduction}

Convolutional Neural Networks(CNNs) have been proven successful in recent years’ application of image recognition. Different from traditional deep learning models, CNNs utilize a unique convolutional and pooling layer structure. With repetitions of this convolutional and pooling layer structure, CNNs are able to extract the features and patterns from the input training data, thus making reliable predictions.\\
Quantum CNNs(QCNNs) have recently emerged as a novel generalization to CNNs\cite{Hur2021}. QCNNs leverage the inherent parallelism of quantum mechanics to perform specific tasks more efficiently than CNNs, and they have the potential to outperform CNNs in some applications. In order to study further the mechanism of QCNNs, this research implemented CNNs and QCNNs, using the same training data and training parameters. Data augmentation techniques in the training process encompass several ways to generate new images that preserve the features and contents according to the existing data. After applying data augmentation in the training process, CNN models tend to learn semantic content better than without data augmentation. As a result, data augmentation helps prevent overfitting and improves the performance of CNN models.\\
By comparison of CNNs and QCNNs after applying data augmentation, the findings of this research provide a better understanding of quantum mechanics in the field of machine learning and its potential applications to improve current models and propose new quantum machine learning algorithms.
\section{Literature Review}
The history of computer vision development can be traced back to the 1970s when researchers started to experiment with approaches to automatically extract image contents and analyse the semantics. Some early researchers\cite{Baumgat1975} used vertex and edge extraction and template matching to detect certain shapes and features in an image. These approaches were able to accomplish some of the tasks but largely relied on manual feature engineering and had limitations in dealing with variations of feature types and scales. \\
A remarkable innovation in computer vision technique came with the development of deep learning in image recognition, especially the emergence of Convolutional Neural Networks. CNNs brought the revolutionary technique by automatically extracting hierarchical features from image data and using this pattern to make predictions on image recognition. A comprehensive survey\cite{Zahangir} investigated the machine learning models regarding deep learning and image recognition proposed from 2012 to 2018, covering AlexNet, VGG-16, GoogleNet-19, ResNet-152, etc., and pointed out that the performance of these models has improved significantly as researchers continue to advance their understanding of deep learning. Ever since 2012 when AlexNet was proposed, more and more convolutional neural networks were invented to improve the performance of image recognition. VGGNet applied deep learning networks with a finer size of filters, GoogleNet introduced parallel computations, and ResNet deliberately neglected some of the nodes to alleviate the vanishing gradient problem. \\
To solve the problem of training data shortage, data augmentation is a technique commonly used in CNNs to increase the size of the training set by generating new data from the existing data through transformations such as rotations, translations, and flips\cite{HAN201843}. When the training data is limited, DA can potentially increase the number of training images and improve the performance of CNNs. This method is able to produce numerous generated images that are similar to the original sample and helps prevent overfitting problems in the training process, especially when the number of training data is small. It is also concluded that data augmentation could improve CNNs performance and geometric data augmentation outperformed photometric data augmentation in terms of Top-1 and Top-5 scores\cite{Taylor2018}.\\
As a promising alternative, QCNNs newly emerged in recent years\cite{Cong2019}, utilizing the concepts of quantum mechanics. As a combination of quantum computing and deep learning, QCNNs leverage the mechanism of quantum superposition and quantum entanglement to transform the data of an image into quantum bits, also known as qubits, and apply the quantum circuit to conduct CNNs-like computations, realising binary classification. \\
Similar to CNNs, the first step of image data processing in QCNNs is to transform the input data into another space. A quantum feature transformation is normally conducted by mapping the input data $X$ to $X\;'$, in a Hilbert Space\cite{Schuld2019}. There are several methods of data mapping, including amplitude embedding, qubit embedding, dense qubit embedding and hybrid embedding. Amplitude embedding transforms the input data of size $N$ into the amplitudes of an $n$ qubit quantum state, where $N = 2^n$. Mathematically, the complexity of the original data input is reduced exponentially to $O(log_2N)$, achieving a remarkable advantage in training efficiency\cite{Araujo2021}. Qubit embedding is similar to classic feature mapping, where $N$ number of data input will be embedded into $N$ number of qubits. Each of the embedded classic data $x_i$ will be re-scaled from 0 to $\pi$ so that the qubit can use the form of: 
\begin{equation}
\Ket{\psi} = cos\frac{x_i}{2}\Ket{0}+sin\frac{x_i}{2}\Ket{1}
\end{equation}
Dense qubit embedding is an optimization of qubit embedding, which only embeds $N$ number of data input to $\frac{N}{2}$ number of qubits\cite{LaRose2020}. Hybrid embedding is a combination of both amplitude embedding and qubit embedding.\\
The convolutional and pooling layers in QCNNs are made of various parameters of unitary gates, including one-qubit gates and two-qubit gates\cite{Rajesh2021}. For example, a convolutional layer of QCNNs contains unitary gates for each qubit and Ising gates between neighbouring qubits, and each of the unitary gates will rotate the qubit at a certain angle around a specific axis. The pooling layer involves controlled measurement gates. After the convolutional layer, half of the qubits will be measured and then collapse. According to the measured result, a specialized unitary gate will be applied to the rest qubits. By doing the repetitions of convolutional and pooling layers circuit, the number of qubits will decrease exponentially, until a very small number of qubits is left, normally around 2 or 3 qubits. The classification results are revealed by measuring one of the remaining qubits in the quantum system. Due to the nature of quantum mechanics, a qubit will only be in state $\Ket{0}$ or $\Ket{1}$ after measurement, QCNNs circuit can realise a binary classification.\\
Despite their potential, QCNNs are still in the early stage of development, and there is a need for further research to explore their capabilities and limitations. For example, given the limitation of training data, it is worth studying whether QCNNs could incorporate a similar process of data augmentation to enhance image recognition performance.
\section{Methodology}
To study quantum mechanics and quantum machine learning, throughout this research, the performance of CNNs and QCNNs are investigated and compared, together with the implementation of data augmentation in training process.\\
In this research, the three training datasets for both CNNs and QCNNs are $8\times8$ MNIST hand-written digits images imported through the sklearn library package, $28\times28$ Fashion MNIST images from Kaggle and a more complex $32\times32$ cat/dog dataset from Kaggle. The first part of the research is to implement the CNNs and data augmentation methods. The second part is the realization of QCNNs and data augmentation methods.\\

\subsection{Classical CNNs}

CNNs mainly contain a structure of convolutional layers, pooling layers, flattened layers, and activation functions. According to the size of the training data input, the detailed structure and hyperparameters may vary largely.\\
The activation function for convolutional layers is chosen to be Rectified Linear Unit (ReLU), introducing non-linearity to the neural network. The pooling method in the pooling layer is to extract the maximum value of each $2\times2$ window from the feature map passed from the last convolutional layer, with a stride parameter of two. In this design, the pooling window will move with a step size of 2 in both the horizontal and vertical directions.\\
After going through the repetitions of convolutional and pooling layers designed above, the multidimensional input tensor will be flattened to a 1D array. The purpose of this layer is to transform the higher dimensional information into a vector, which will be passed to a fully connected dense layer for the next activation function. In a dense layer, each of the neurons is connected to the other neurons in the next layer. This structure will preserve the captured features at the maximum level. The activation function in the last layer is SoftMax, the loss function is Sparse Categorical Cross-entropy. The optimization method is the Adam optimizer, which has an adaptive learning rate and efficiency.\\

\subsection{Data Augmentation}

\begin{figure}[H]
\centering
\begin{minipage}{0.25\textwidth}
  \includegraphics[width=\linewidth]{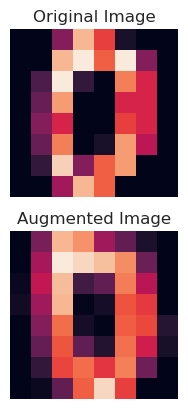}
  \caption{Digit 0}\label{DA img digits}
\end{minipage}\hfill
\begin{minipage}{0.25\textwidth}
  \includegraphics[width=\linewidth]{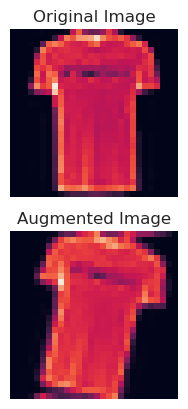}
  \caption{A T-shirt}\label{DA img2}
\end{minipage}\hfill
\begin{minipage}{0.25\textwidth}%
  \includegraphics[width=\linewidth]{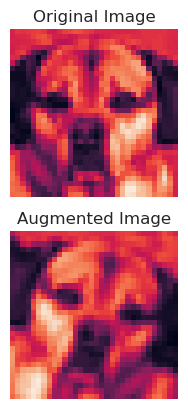}
  \caption{A dog face}\label{DA img}
\end{minipage}
\end{figure}

Data augmentation is a useful tool to improve the performance of CNNs, especially when the training data is in shortage. It applies various random transformations or modifications to the original data on a reasonable scale, keeping the semantic contents in the augmented images the same as the original data. In a way, data augmentation works as a regularization method to prevent overfitting in the training process, by generating diversity and variations in each batch of the training data.\\
There are several commonly adopted data augmentation methods in CNNs\cite{Taylor2018}. We applied the following methods: Random flip is to flip the original image in the horizontal or vertical direction; Random rotation is to rotate the original image by a reasonable angle; Random contrast adjusts the pixel value in the original image to increase or decrease the difference between dark and light colours.\\
The data augmentation method in this research is realised by TensorFlow Keras layers\cite{tensorflow2015-whitepaper}. The random rotation method applies an rotation to the original data by an angle within 0.05 radian. The random contrast method manipulates the contrast of pixel value between dark and light colours between 0.9 and 1.1. Random Flip function flips the image horizontally. We applied random flip and random rotation to Fashion MNIST and cat/dog datasets, random rotation and random contrast to hand-written digits. Figure \ref{DA img digits}, \ref{DA img2}, \ref{DA img} are some examples of the augmented images.\\

\subsection{Quantum CNNs}
\begin{figure}[H]
	\centering
	\includegraphics[scale=0.22]{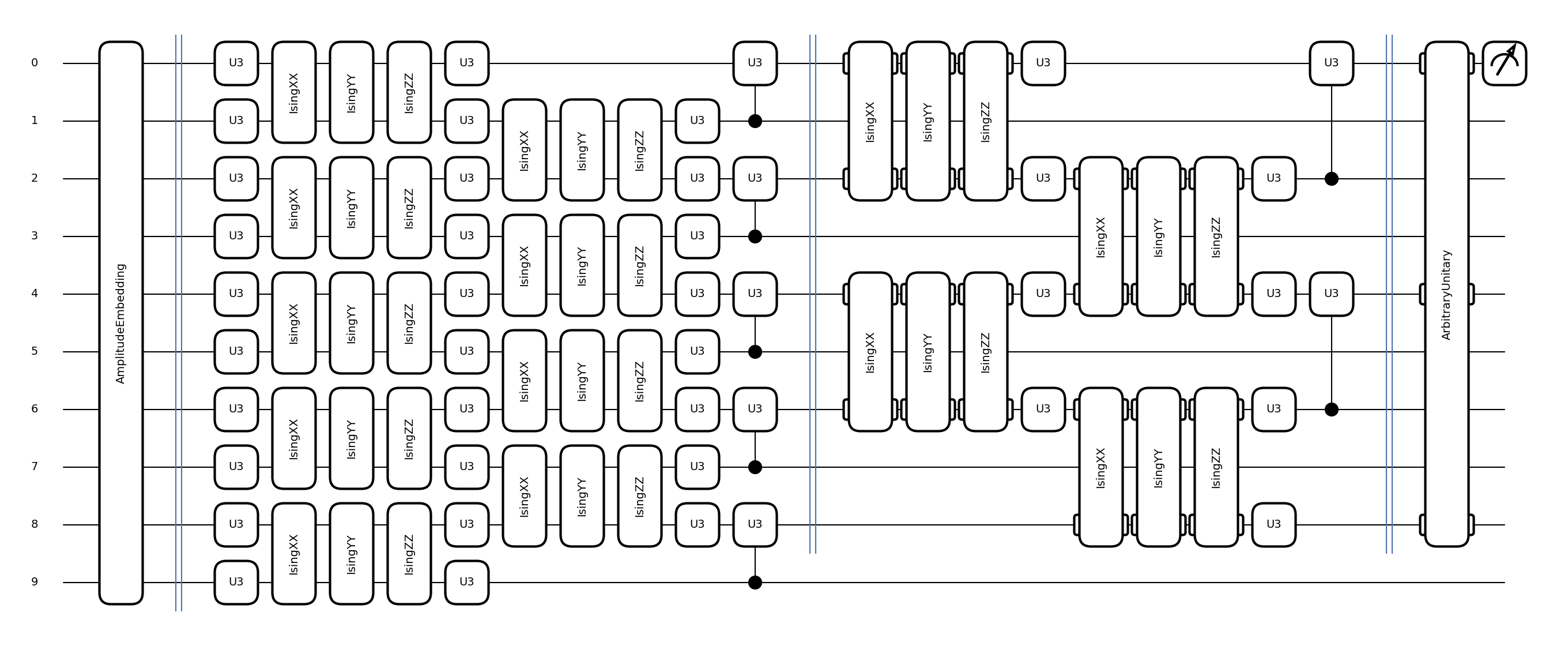}
	\caption{Structure of QCNNs circuit \cite[]{Kottmann2022}}\label{QCNN}
\end{figure}

QCNNs normally need to transform image pixel value matrix into quantum information. Some of the commonly used approaches include amplitude embedding, qubit embedding, dense qubit embedding and hybrid embedding\cite{Choi2020}. Amplitude embedding transforms input data from $N$ pixel to $log_2N$ qubits, which facilitates the computation by reducing the number of input data. Qubit embedding transforms the input data into the same number of qubits, which optimises the depth of the quantum circuit. Amplitude embedding and qubit embedding represents two extremes of the feature mapping approaches, either computation complexity or quantum circuit length. Dense qubit embedding and hybrid embedding are in between these two extremes.\\
Throughout the implementation of QCNNs in this research, amplitude embedding is chosen to transform an input image with $N$ number of pixel value, into a number of $log_2N$ qubits. Due to quantum superposition, each qubit can exist in the state of both $\ket{0}$ and $\ket{1}$ simultaneously. Within a system of $n$ qubits, there are in total $2^n$ possible states. The probability distribution of each state can be seen as the amplitude, thus representing the original pixel value.\\
After mapping the input image into qubits, similar to constructing CNNs structures, we have the convolutional layers and pooling layers circuit. There are some variations of the structures of convolutional layers and pooling layers in QCNNs\cite{Cong2019}\cite{Caro2022}. In this research, we adopted the implementation methods by the Pennylane library tutorial regarding quantum machine learning\cite{Kottmann2022}. \\
In the quantum convolutional layer, each of the qubits is applied with unitary gates and quasilocal unitary gates (Ising gate) associated with the neighbouring qubits, so that they have a certain level of localization. In quantum mechanics, unitary gates are fundamental operators that apply a transformation on a single qubit or group of qubits. For some of the single qubit unitary gates, the Pauli X gate will flip the bit from $\ket{0}$ state to $\ket{1}$ state; the Pauli Z gate will change the qubit phase to the opposite; the Pauli Y gate will change both the state bit and phase; the Hadamard gate will create a superposition of basis state. These one-qubit unitary gates can be all interpreted as applying a rotation on the targeted qubit at a certain angle of $\alpha$ around a specific axis $\hat{n}$. In general, it is represented as:
\begin{equation}
R(\alpha) = exp(-i\frac{\alpha}{2}\hat{n}*\sigma)
\end{equation}
where $\sigma = (X, Y, Z)$ and $\hat{n}*\sigma$ represents the axes of the rotation. For a two-qubit unitary gate, the CNOT gate will apply a Pauli X gate on one qubit if the other qubit is in state $\ket{1}$. \\
In the circuit design for the quantum convolutional layer, we have to realise the similar convolution computation as CNNs do. The kernel swept across the whole image from the left to the right, up to the bottom, realising the computation of local pixel value. In QCNNs circuits, a series of single-qubit and two-qubit gates can achieve similar functions. In the beginning, two unitary gates will be applied to each of the qubits after amplitude embedding. Each of the Unitary gates has three parameters, $\theta,\phi,\lambda$. They correspond to three Euler angles. The following is a matrix representation of the unitary gates:
\begin{equation}
U(\theta,\phi,\lambda) = \begin{bmatrix}
	cos(\frac{\theta}{2}) & -e^{i\lambda}sin(\frac{\theta}{2})\\
	e^{i\phi}sin(\frac{\theta}{2}) & e^{i(\phi+\lambda)}cos(\frac{\theta}{2})
\end{bmatrix}
\end{equation}
After applying unitary gates to the qubits, two-qubit gates of IsingXX, IsingYY and IsingZZ gates are applied to pairs of qubits. The name Ising is derived from the mathematical Ising model that describes the behaviour of magnetic spins. In general, Ising gates realised an interaction function between two qubits in quantum systems. In essence, this circuit forms a convolutional layer for QCNNs.\\
The pooling layer in the quantum circuit usually requires measuring half of the qubits. In quantum theory, when a qubit is measured, it loses its superposition and it collapses. If the measurement is $\ket{1}$, a unitary gate will be applied to the the other qubit. For an $n$-qubit circuit, after applying the first pooling layer, the number of qubits decreases sharply from $n$ to $\frac{n}{2}$.\\
One depth of convolutional and pooling layer is described as above. After several depth of convolutional and pooling layer, the number of qubits left will be normally very little. For example, a 10-qubit circuit with two depth only has three qubits left after the flatten layer. By measuring a remaining qubit, a probability distribution of state $\ket{0}$ and $\ket{1}$ will indicate the binary classification result.
Overall, the structure of a 10-qubit quantum circuit implemented in this research is as Figure \ref{QCNN}.
\noindent
From left to right, the input image data will go through the forward training process and come up with a binary prediction. The back-propagation circuit looks similar to the forward training circuit, but it runs in the opposite direction. Another difference lies in the pooling layers. In forward training, the pooling layer measures half of the qubits, causing quantum collapse, and leaving only half of the qubits. Therefore, in back-propagation, in the pooling layer, it is necessary to add more qubits in place of the collapsed qubits. Running from right to left of the current quantum circuit, the measurement operation needs to be replaced by a qubit. \\
The back-propagation circuit is also known as the Multiscale Entanglement Renormalization Ansatz (MERA). It is a well-known tensor network ansatz from the field of quantum many-body problems. MERA proposes an approach to represent and relate to quantum many-body systems in one-dimensional space. It can provide a finer representation of the quantum state by systematically extracting the high-frequency entanglement.\\
In the training process, the loss function adopted is the mean squared error (MSE), the optimizer is the Adam optimizer, and the learning rate is initially set to be 0.1, decaying by 5\% as the epochs increase.

\section{Results}

\subsection{CNNs and Data Augmentation}

\begin{figure}[H]
	\centering
	\includegraphics[scale=0.44]{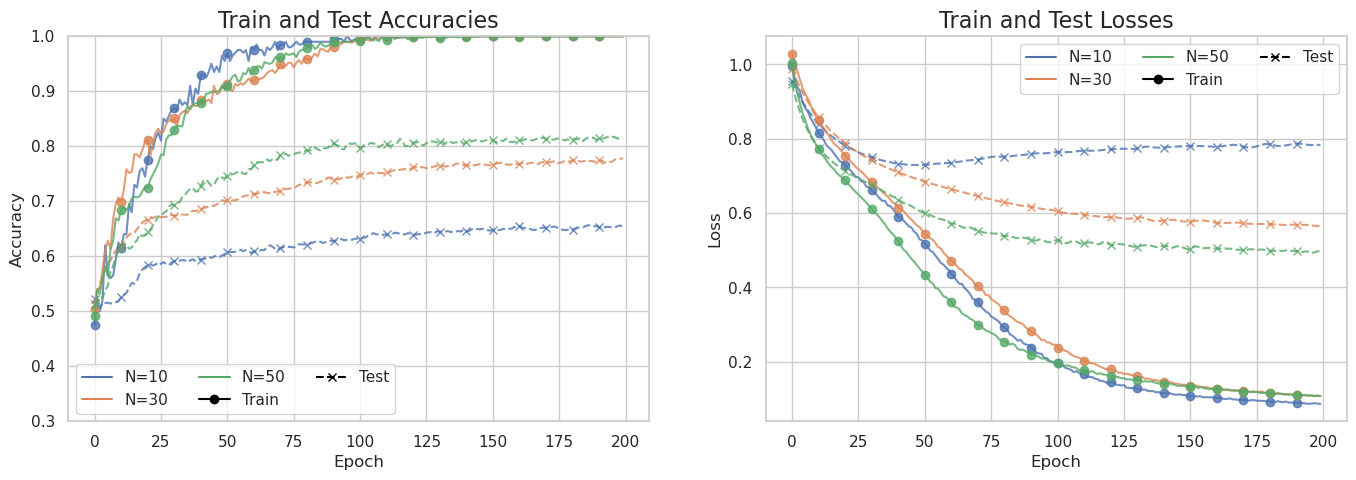}
	\caption{Cat/Dog CNNs performance}
        \label{cat/dog CNNs performance}
\end{figure}

In the experiments of CNNs, all the $8\times8$ hand-written digits, $28\times28$ Fashion MNIST and $32\times32$ cat/dog datasets are applied. The three datasets show the same results that data augmentation can improve CNNs performance. In this section, the results presented are from cat/dog datasets. The number of training data N is set to 10, 30, 50, and the number of testing data is the same amount of 100. 200 epochs are enough for the CNNs model to converge. Each model was trained 20 repetitions, and the results are based on the average the 20 repetitions. The results of cat/dog datasets training and testing accuracy and losses are shown as Figure \ref{cat/dog CNNs performance}.\\

\begin{figure}[H]
	\centering
	\includegraphics[scale=0.44]{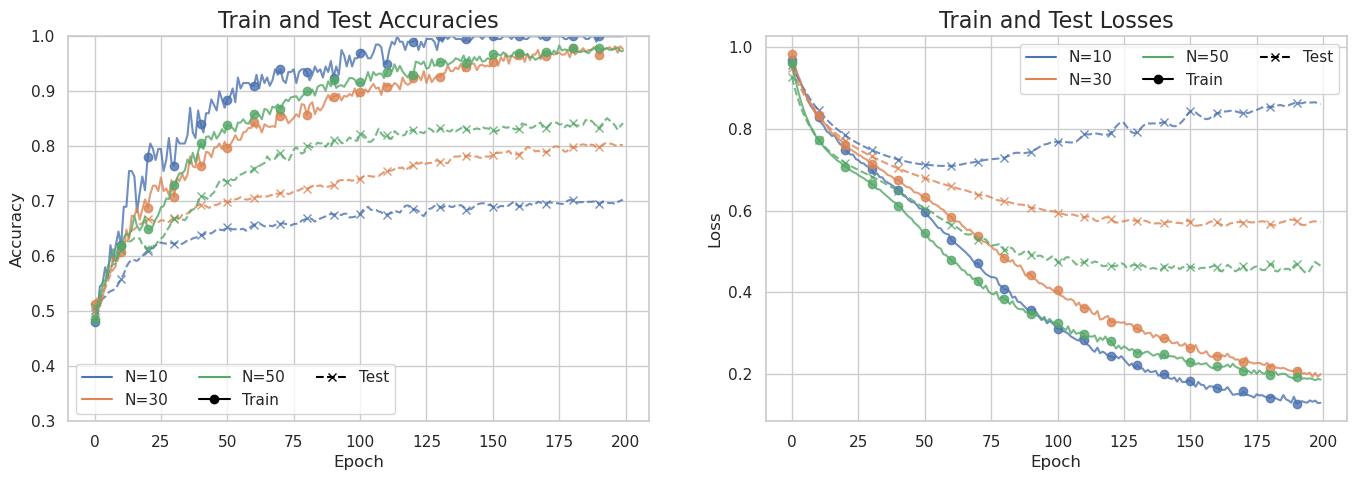}
	\caption{Cat/Dog CNNs performance with data augmentation}
        \label{cat/dog CNNs performance with data augmentation}
\end{figure}

\noindent
From the figures we can observe that, as the epoch increased, the testing accuracy for N equals 10, 30 and 50 increased gradually to around 0.66, 0.78 and 0.81 respectively. The testing losses dropped for 30 and 50 training data, but fluctuated below 0.8 for 10 training data. This observation means that the CNNs model tends to overfit when the training data is sparse, causing a low prediction accuracy, which is not preferable for any machine learning models.\\
In order to verify that data augmentation can improve the performance of CNNs, the random rotation layers and random flip layers are implemented to the original cat/dog image before the convolutional layers. The number of epochs is 200 when the CNNs model converges after applying data augmentation. Figure \ref{cat/dog CNNs performance with data augmentation} are the training and testing accuracy and losses:\\

\noindent

After applying data augmentation, the testing accuracy of N equals 10, 30 and 50 increased to above 0.7, 0.8 and 0.83. This performance is higher than before data augmentation. The testing losses for 30 and 50 training data are also lower, although it is slightly higher for 10 training data.\\
In comparison to CNNs, the testing accuracy is relatively higher, and the losses are smaller, especially when the training data is insufficient like N equals 10 and 30. This result verified that data augmentation can improve the performance of the CNNs model.\\

\subsection{QCNNs and Data Augmentation}
In comparison of QCNNs before and after the application of data augmentation, it is unexpectedly observed that the performance didn’t improve at all. In order to collect more data to compare the impact of data augmentation on QCNNs, we implemented the binary classification between 0 and all the other hand-written digits from 1 to 9, as shown in Table \ref{Accuracy of Quantum CNNs}, \ref{Accuracy of Quantum CNNs after DA}.\\
\begin{table}[H]
\centering
\begin{tabular}{c c c c c c c c c c c}
 \hline
 \multicolumn{10}{c}{Accuracy of hand-written digit classification} \\
 No. of input&1&2&3&4&5&6&7&8&9 \\
 \hline
 5   &82.8	&78.3	&79.2	&77.5	&73.7	&77.1	&80.9	&74.9	&69.8 \\ 
 
 10	&91.1	&88.1	&85.6	&85.0	&77.9	&81.9	&89.3	&78.2	&75.1\\
 
 30	&97.8	&96.3	&97.5	&97.8	&91.4	&91.8	&96.2	&90.7	&87.3\\
 
 50	&98.2&	98.9	&97.7	&98.4	&96.1	&95.7	&99.9	&97.3	&91.1\\
 \hline
\end{tabular}
\caption{Accuracy of QCNNs}
\label{Accuracy of Quantum CNNs}
\end{table}

\begin{figure}[H]
	\centering
	\includegraphics[scale=0.44]{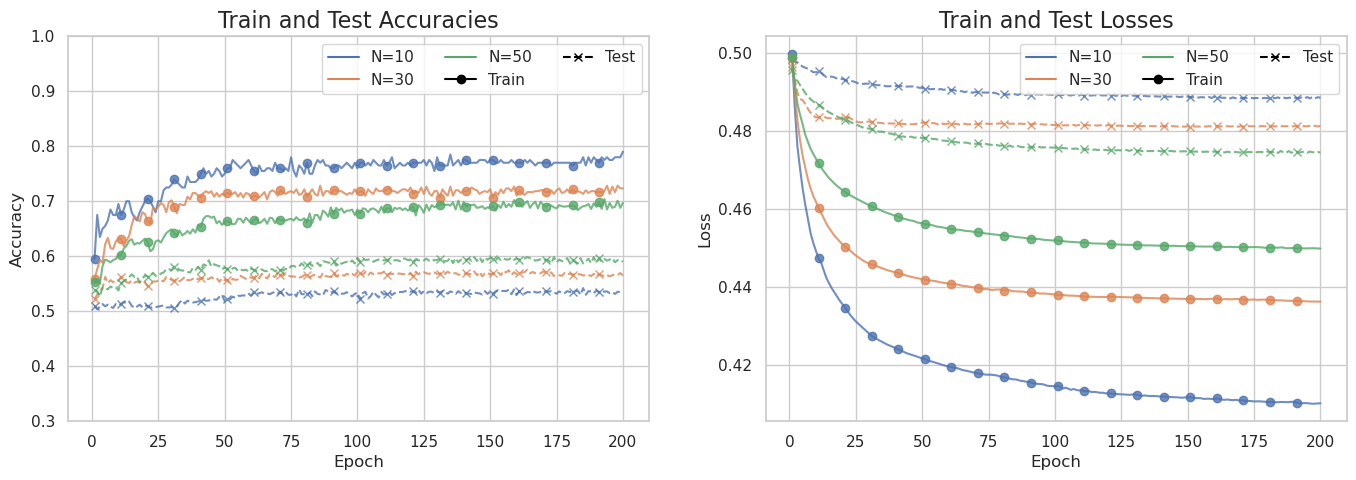}
	\caption{Cat/Dog QCNNs performance}
        \label{cat/dog QCNNs performance}
\end{figure}

\begin{table}[H]
\centering
\begin{tabular}{c c c c c c c c c c c}
 \hline
 \multicolumn{10}{c}{Accuracy of hand-written digit classification after DA} \\
 
 No. of input&1&2&3&4&5&6&7&8&9 \\
 \hline
 5   &82.4	&77.2	&77.4	&74.6	&71.3	&74.2	&78.1	&71.3	&67.4 \\ 
 
 10	&88.3	&86.4	&83.5	&82.9	&75.1	&78.6	&88.8	&80.7	&74.2\\
 
 30	&95.2	&95.7	&95.0	&94.0	&89.2	&91.2	&97.8	&92.4	&82.6\\
 
 50	&97.8	&97.0	&98.2	&96.8	&91.4	&96.7	&99.3	&95.0	&91.1\\
 \hline
\end{tabular}
\caption{Accuracy of QCNNs after DA}
\label{Accuracy of Quantum CNNs after DA}
\end{table}
It is clear that the accuracy of QCNNs, implemented with data augmentation, is slightly worse than before. This observation reveals an interesting discovery that data augmentation indeed cannot improve the performance of QCNNs models.\\
The hand-written digits data is just in $8\times8$ pixel size, so that the QCNNs circuit only contains 6 qubits to transform the 64 pixel value. In order to expand our experiments, we used the cat/dog dataset, originally in $512 \times 512$ size, and compressed it to $32 \times 32$. So the quantum circuit contains 10 qubits. The pixel value increased from 64 to 1024, requiring a more complex circuit. The QCNNs performance on this cat/dog dataset is presented in figure \ref{cat/dog QCNNs performance}.\\

\begin{figure}[H]
	\centering
	\includegraphics[scale=0.44]{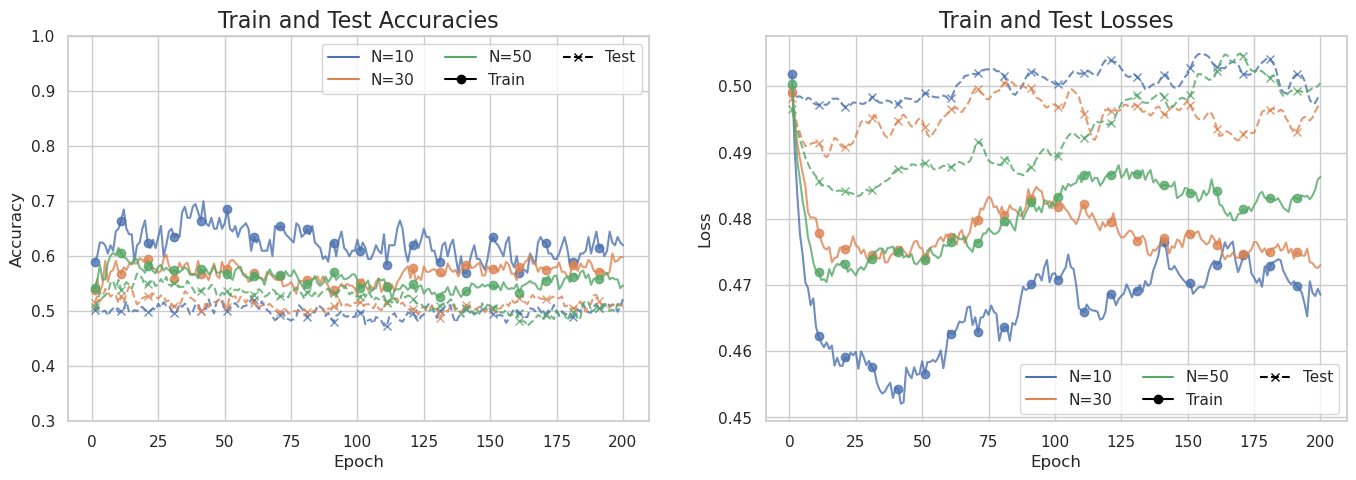}
	\caption{Cat/Dog QCNNs performance with data augmentation}
        \label{cat/dog QCNNs performance with data augmentation}
\end{figure}

\begin{figure}[H]
	\centering
	\includegraphics[scale=0.7]{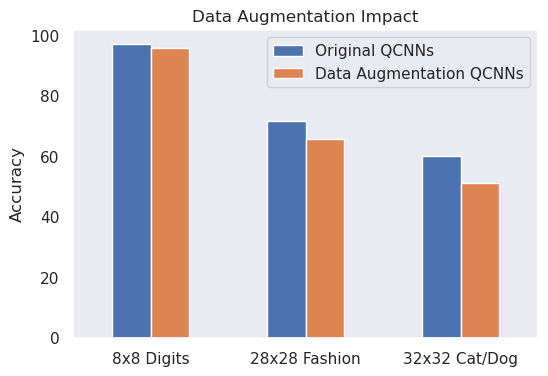}
	\caption{Impact of data augmentation for three different datasets}
        \label{Impact of data augmentation for three different datasets}
\end{figure}

\noindent
As the epochs progress, the training accuracy demonstrates an upward trend; however, the testing performance remains marginally above the baseline, ranging from 0.55 to 0.6 for training image numbers N between 10 and 50. The contrast between high training accuracy and low testing accuracy underscores the suboptimal performance of the QCNNs structure in tackling complex classification problems, particularly when confronted with a scarcity of training data.\\
After adopting data augmentation, the performance is shown as Figure \ref{cat/dog QCNNs performance with data augmentation}\\
\noindent
The training and testing accuracy fluctuate around the baseline, indicating that not only has data augmentation failed to enhance QCNNs, but it has also led to a potential decline in performance. This observation aligns with the performance results observed in QCNNs when applied to hand-written digits.\\
Furthermore, the $28\times28$ size Fashin-MNIST data was also adopted for this QCNNs structure. The comparison of different datasets is shown in this bar chart \ref{Impact of data augmentation for three different datasets}
\noindent

As the dimensions of the images grow, the intricacy of image classification also rises. Consequently, the effectiveness of QCNNs diminishes, and the efficacy of data augmentation further deteriorates. The quantum circuit structure exhibits limitations in scalability and struggles to adeptly accommodate both intricate and augmented datasets.\\

\section{Discussion}
According to the simulation results in the last section, it is verified that data augmentation can improve the performance of CNNs models in the situation of data shortage during the learning process. We also discovered that data augmentation couldn’t improve the performance of QCNN models in most situations. It is worth discussing factors why QCNN models are insensitive to data augmentation.\\
The first reason can be attributed to the nature of mapping an image into quantum information during the feature transformation. We adopt amplitude embedding, which maps a number of $N$ values into the amplitudes of possible states in a system containing $log_2N$ qubits. In this process, the number of input values decreased exponentially. Although the newly generated images by data augmentation are slightly different from the original data, after applying amplitude embedding, the qubits we get will be in a position far different from the qubits transformed by the original images. If the position of one qubit is changed, the amplitudes of the whole system will change. After applying data augmentation and then amplitude embedding, training the QCNNs model with this huge difference from the authentic data, it is reasonable that the accuracy of classification can be worse. In order to solve this issue, hypothetically, an innovative approach of data embedding could be designed to mitigate these deviations from the original data, keeping the slight differences with exponentially improved computation complexity.\\
Another reason why data augmentation cannot improve the performance of QCNNs may lie in the structure of pooling layers. In one layer of the pooling function, half of the qubits will be measured, leading to inevitable state collapses, and the number of the qubits will be reduced by half. This process causes the number of amplitudes to plunge from $2^n$ to $2^\frac{n}{2}$. In the experiment of cat/dog datasets, 1024 values drop to 32 values after applying one round of convolutional and pooling layers. This dramatic reduction in the number of values naturally prevents overfitting of the QCNNs model. But in a complex classification problem, the pooling layers may result in an under-trained model that only performs slightly better than the baseline.\\
Apart from these observations due to data augmentation, it is also noted that this current QCNNs circuit structure is lacking scalability. For the $8\times8$ hand-written digits dataset, QCNN classification performance is slightly lower than CNNs. However, for the $32\times32$ cat/dog dataset, QCNNs performance is slightly higher than the baseline, whereas CNN's performance could reach 0.85 accuracy. This comparison revealed that the current convolutional and pooling circuit structure couldn't learn the differences in complex features. This is partly because CNNs apply convolutional and pooling layers to 2D features before the final flatten layer, but QCNNs need to transform 2D features into 1D qubit array using amplitude embedding before convolutional and pooling circuit. This structure limits the QCNNs circuit to learn features only from one dimensional computations.\\
Apart from that, the number of parameters in one depth of the convolutional and pooling layer is quite small. Because of the shared parameter theory\cite{Caro2022}, there are only 18 parameters in each depth. The total number of parameters $P$ of an arbitrary QCNNs circuit is:
\begin{equation}
P(d,r) = 18 \times d + (4^r - 1)
,\end{equation}
where $d$ is the number of depth and $r$ is the number of remaining qubits in the last flatten layer. The total number of parameters for the QCNNs circuit on the cat/dog dataset is merely $18 \times 2 + (4^3-1) = 99$. According to another research\cite{Larocca}, a 10-qubit Variational quantum eigensolver (VQE) needs at least 150 parameters to perform classification successfully. It is reasonable that QCNNs with this little number of parameters cannot perform well on large scale data.\\ 
Overall, QCNNs differ significantly from CNNs in terms of mathematical computations. Today, quantum machine learning is still a newly emerging technology, and we don’t know everything about its capabilities and limitations. There may be additional reasons why QCNNs models are insensitive to data augmentation. Exploring this topic further through research would be worthwhile.

\section{Conclusion}
In this research, we implemented CNN and QCNN  models and applied them to three different commonly used datasets. Afterwards, data augmentation methods, including random rotation, random flip and random contrast, were applied to CNN and QCNN models, respectively. The results verified that data augmentation can improve the prediction accuracy of the CNNs model, especially when the training images are in shortage. As for the QCNNs model, it is surprisingly observed that the binary classification outcome did not improve after adopting data augmentation methods. In most cases, data augmentation decreases QCNN classification accuracy.\\
This research investigated innovative techniques in quantum mechanics and quantum machine learning and compared the difference between CNN and QCNN in terms of the impact of data augmentation. The findings of this research leveraged the inherent characteristics of QCNN and provided a deeper understanding of the mechanisms behind quantum machine learning.

\section{Future Work}
As an initial study into the quantum machine learning algorithm, this research focused on a specific QCNNs model\cite{Cong2019}. In the future, there are some other work we can do to understand it better.\\
An approach is to improve the natural binary classification model to multi-class classification. In some other quantum algorithms, for example, Shor’s algorithm\cite{Monz2015}, the solution is acquired by measuring multiple qubits in the last layer. By transforming the outcome into a 10-base number, the solutions can be more than just 0 and 1. A similar mechanism could be adopted for QCNNs models. At the last flattened layer, there is normally more than one qubit left, for example, $n$ qubits. If we can measure all the remaining qubits and acquire a probability distribution of $2^n$ values, it is possible to realise a multi-class classification.\\
In the future, it will be feasible to realise the QCNN models on a real quantum computing device. Currently, we have implemented the QCNN circuits on a classical computer using Python environment simulations. One of the most significant reasons is the inevitable noise caused by real quantum devices. During the training process of the QCNN models, the quantum circuit will be executed repeatedly, often with thousands of iterations. When working on real quantum devices, each of the gates in the quantum circuit carries a low probability of inducing errors. These errors can significantly disrupt the forward training and backpropagation processes, resulting in failure or degraded performance. Nowadays, it is possible to prevent these errors by applying quantum error correction codes, for example, the surface code\cite{Fowler2008}.\\
Overall, the field of quantum machine learning models encompasses a wide range of potential areas worth exploring. These models' unique and extraordinary nature opens up numerous possibilities for further research and investigation. The intricate interplay between quantum principles and neural networks presents an exciting frontier for interdisciplinary collaboration and innovation. With each step forward, we will better understand the power and capabilities of quantum machine learning.\\
\\
\printbibliography


\end{document}